\documentstyle[11pt,aaspp4]{article}  

\begin{document}

\title{Extreme Kuiper Belt Object 2001 QG$_{298}$ and the Fraction of Contact Binaries}  
\author{Scott S. Sheppard and David Jewitt\footnote{Visiting Astronomer, W.M. Keck Observatory, which operates as a scientific partnership among the California Institute of Technology, the University of California, and the National Aeronautics and Space Administration.  The Observatory was made possible by the generous financial support of the W.M. Keck Foundation}}    
\affil{Institute for Astronomy, University of Hawaii, \\
2680 Woodlawn Drive, Honolulu, HI 96822 \\ sheppard@ifa.hawaii.edu, jewitt@ifa.hawaii.edu}

\begin{abstract}  
Extensive time-resolved observations of Kuiper Belt object 2001
QG$_{298}$ show a lightcurve with a peak-to-peak variation of $1.14
\pm 0.04$ magnitudes and single-peaked period of $6.8872 \pm 0.0002$
hr.  The mean absolute magnitude is 6.85 magnitudes which corresponds
to a mean effective radius of 122 (77) km if an albedo of 0.04 (0.10)
is assumed. This is the first known Kuiper Belt object and only the
third minor planet with a radius $>$ 25 km to display a lightcurve
with a range in excess of 1 magnitude.  We find the colors to be
typical for a Kuiper Belt object ($B-V = 1.00\pm 0.04$, $V-R = 0.60\pm
0.02$) with no variation in color between minimum and maximum light.
The large light variation, relatively long double-peaked period and
absence of rotational color change argue against explanations due to
albedo markings or elongation due to high angular momentum.  Instead,
we suggest that 2001 QG$_{298}$ may be a very close or contact binary
similar in structure to what has been independently proposed for the
Trojan asteroid 624 Hektor.  If so, its rotational period would be
twice the lightcurve period or $13.7744 \pm 0.0004$ hr.  By correcting
for the effects of projection, we estimate that the fraction of
similar objects in the Kuiper Belt is at least $\sim$10\% to 20\% with
the true fraction probably much higher.  A high abundance of close and
contact binaries is expected in some scenarios for the evolution of
binary Kuiper Belt objects.

\end{abstract}


\section{Introduction}
The Kuiper Belt is a long-lived region of the Solar System just beyond
Neptune where the planetisimals have not coalesced into a planet.  It
contains about 80,000 objects with radii greater than 50 km (Trujillo,
Jewitt \& Luu 2001) which have been collisionally processed and
gravitationally perturbed throughout the age of the Solar System.  The
short-period comets and Centaurs are believed to originate from the
Kuiper Belt (Fernandez 1980; Duncan, Quinn \& Tremaine 1988).

Physically, the Kuiper Belt objects (KBOs) show a large diversity of
colors from slightly blue to ultra red ($V-R \sim 0.3$ to $V-R \sim
0.8$, Luu and Jewitt 1996) and may show correlations between colors,
inclination and/or perihelion distance (Jewitt \& Luu 2001; Trujillo
\& Brown 2002; Doressoundiram et al. 2002; Tegler \& Romanishin 2003).
Spectra of KBOs are mostly featureless with a few showing hints of
water ice (Brown, Cruikshank \& Pendleton 1999; Jewitt \& Luu 2001;
Lazzarin et al. 2003).  The range of KBO geometric albedos is still
poorly sampled but the larger ones likely have values between 0.04 to
0.10 (Jewitt, Aussel \& Evans 2001; Altenhoff, Bertoldi \& Menten
2004).  Time-resolved observations of KBOs show that $\sim 32 \%$ vary
by $\ge 0.15$ magnitudes, $18 \%$ by $\ge 0.40$ magnitudes and $12 \%$
by $\ge 0.60$ magnitudes (Sheppard \& Jewitt 2002; Ortiz et al. 2003;
Lacerda \& Luu 2003; Sheppard \& Jewitt 2004). One object, (20000)
Varuna, displays a large photometric range and fast rotation which is
best interpreted as a structurally weak object elongated by its own
rotational angular momentum (Jewitt \& Sheppard 2002).  A significant
fraction of KBOs appear to be more elongated than main-belt asteroids
of similar size (Sheppard \& Jewitt 2002). The KBO phase functions are
steep, with a median of $0.16$ magnitudes per degree between phase
angles of 0 and 2 degrees (Sheppard \& Jewitt 2002; Schaefer \&
Rabinowitz 2002; Sheppard \& Jewitt 2004).

About $4\% \pm 2\%$ of the KBOs are binaries with separations $\ge
0.15 \arcsec$ (Noll et al. 2002) while binaries with separations $\ge
0.1 \arcsec$ may constitute about $15 \%$ of the population (Trujillo
2003, private communication).  All the binary KBOs found to date
appear to have mass ratios near unity, though this may be an
observational selection effect.  The mechanism responsible for
creating KBO binaries is not clear.  Formation through collisions is
unlikely (Stern 2002).  Weidenschilling (2002) has proposed formation
of such binaries through complex three-body interactions which would
only occur efficiently in a much higher population of large KBOs than
can currently be accounted for.  Goldreich, Lithwick \& Sari (2002)
have proposed that KBO binaries could have formed when two bodies
approach each other and energy is extracted either by dynamical
friction from the surrounding sea of smaller KBOs or by a close third
body.  This process also requires that the density of KBOs was
$\sim10^2$ to $10^3$ times greater than now.  They predict that closer
binaries should be more abundant in the Kuiper Belt while
Weidenschilling's mechanism predicts the opposite.

The present paper is the fourth in a series resulting from the Hawaii
Kuiper Belt variability project (HKBVP, see Jewitt \& Sheppard 2002;
Sheppard \& Jewitt 2002; Sheppard \& Jewitt 2004).  The practical aim
of the project is to determine the rotational characteristics
(principally period and shape) of bright KBOs ($m_{R} \le 22$) in
order to learn about the distributions of rotation period and shape in
these objects.  In the course of this survey we found that 2001
QG$_{298}$ had an extremely large light variation and a relatively
long period.  We have obtained optical observations of 2001 QG$_{298}$
in order to accurately determine the rotational lightcurve and
constrain its possible causes.  2001 QG$_{298}$ has a typical Plutino
orbit in 3:2 mean-motion resonance with Neptune, semi-major axis at
39.2 AU, eccentricity of 0.19 and inclination of 6.5 degrees.

\section{Observations}

We used the University of Hawaii (UH) 2.2 m diameter telescope atop
Mauna Kea in Hawaii to obtain R-band observations of 2001 QG$_{298}$
on three separate observing runs each covering several nights: UT
September 12 and 13 2002; August 22, 26, 27 and 28 2003; September 27,
28 and 30 2003.  Two different CCD cameras were employed.  For the
September 2002 and September 2003 observations we used a $2048 \times
2048$ pixel Tektronix CCD ($24$ $\micron$ pixels) camera with a
$0.\arcsec 219$ pixel$^{-1}$ scale at the f/10 Cassegrain focus.  An
antireflection coating on the CCD gave very high average quantum
efficiency (0.90) in the R-band.  The field-of-view was $7\arcmin .5
\times 7\arcmin .5$.  For the August 2003 observations we used the
Orthogonal Parallel Transfer Imaging Camera (OPTIC).  OPTIC has two
$4104 \times 2048$ pixel Lincoln Lab CCID28 Orthogonal Transfer CCDs
developed to compensate for real-time image motion by moving the
charge on the chips to compensate for seeing variations (Tonry, Burke
\& Schechter 1997).  Howell et al. (2003) have demonstrated that these
chips are photometrically accurate and provide routine sharpening of
the image point spread function.  There is a $\sim15 \arcsec$ gap
between the chips.  The total field-of-view was $9\arcmin .5 \times
9\arcmin .5$ with $15$ $\micron$ pixels which corresponds to $0.14
\arcsec$ pixel$^{-1}$ scale at the f/10 Cassegrain focus.  The same
R-band filter based on the Johnson-Kron-Cousins photometric system was
used for all UH 2.2 m observations.

In addition we used the Keck I 10 m telescope to obtain BVR colors of
2001 QG298 at its maximum and minimum light on UT August 30, 2003.
The LRIS camera with its Tektronix $2048 \times 2048$ pixel CCD and
$24$ $\micron$ pixels (image scale $0.\arcsec 215$ pixel$^{-1}$) was
used (Oke et al. 1995) with the facility broadband BVR filter set.
Due to a technical problem with the blue camera side we used only the
red side for photometry at BV and R.  The blue filter response was cut
by the use of a dichroic at 0.460 $\micron$.

All exposures were taken in a consistent manner with the telescope
autoguided on bright nearby stars.  The seeing ranged from $0.\arcsec
6$ to $1.0\arcsec $ during the various observations.  2001 QG$_{298}$
moved relative to the fixed stars at a maximum of $3\arcsec .5$
hr$^{-1}$ corresponding to trail lengths $\leq$ $0.\arcsec 43 $ in the
longest (450 sec) exposures.  Thus motion of the object was
insignificant compared to the seeing.

Images from the UH telescope were bias-subtracted and then
flat-fielded using the median of a set of dithered images of the
twilight sky.  Data from Keck were bias subtracted and flattened using
flat fields obtained from an illuminated spot inside the closed dome.
Landolt (1992) standard stars were employed for the absolute
photometric calibration.  To optimize the signal-to-noise ratio we
performed aperture correction photometry by using a small aperture on
2001 QG$_{298}$ ($0.\arcsec 65$ to $0.\arcsec 88$ in radius) and both
the same small aperture and a large aperture ($2.\arcsec 40$ to
$3. \arcsec 29$ in radius) on (four or more) nearby bright field
stars.  We corrected the magnitude within the small aperture used for
the KBOs by determining the correction from the small to the large
aperture using the field stars (c.f. Tegler and Romanishin 2000;
Jewitt \& Luu 2001; Sheppard \& Jewitt 2002).  Since 2001 QG$_{298}$
moved slowly we were able to use the same field stars from night to
night within each observing run, resulting in very stable relative
photometric calibration from night to night.  The observational
geometry for 2001 QG$_{298}$ on each night of observation is shown in
Table~1.

\section{Results}

Tables~2 and 3 show the photometric results for 2001 QG$_{298}$.  We
used the phase dispersion minimization (PDM) method (Stellingwerf
1978) to search for periodicity in the data.  In PDM, the metric is
the so-called $\Theta$ parameter, which is essentially the variance of
the unphased data divided by the variance of the data when phased by a
given period.  The best-fit period should have a very small dispersion
compared to the unphased data and thus $\Theta <<$ 1 indicates that a
good fit has been found.

2001 QG$_{298}$ showed substantial variability ($\sim 1.1$ magnitudes
with a single-peaked period near $6.9$ hr) in R-band observations from
two nights in September 2002.  We obtained further observations of the
object in 2003 to determine the lightcurve with greater accuracy.  PDM
analysis of all the apparent magnitude R-band data from the September
2002 and August and September 2003 observations shows that 2001
QG$_{298}$ has strong $\Theta$ minima near the periods $P = 6.88$ hr
and $P = 13.77$ hr, with weaker alias periods flanking these
(Figure~\ref{fig:pdmqg}).  We corrected the apparent magnitude data
for the minor phase angle effects (we used the nominal 0.16 magnitudes
per degree found in Sheppard \& Jewitt 2003) and light travel-time
differences of the observations to correspond to the August 30, 2003
observations.  We then phased the data to all the peaks with $\Theta <
0.4$ and found only the 6.8872 and 13.7744 hour periods to be
consistent with all the data (Figures~\ref{fig:phaseqgsingle} and
\ref{fig:phaseqgdouble}).  Through a closer look at the PDM plot
(Figure~\ref{fig:pdmqgsmall}) and phasing the data we find best fit
periods $P = 6.8872 \pm 0.0002$ hr (a lightcurve with a single maximum
per period) and $P = 13.7744 \pm 0.0004$ hr (two maxima per period as
expected for rotational modulation caused by an aspherical shape).
The double-peaked lightcurve appears to be the best fit with the
minima different by about 0.1 magnitudes while the maxima appear to be
of similar brightness.  The photometric range of the lightcurve is
$\Delta m = 1.14 \pm 0.04$ magnitudes.

The Keck BVR colors of 2001 QG$_{298}$ show no variation from minimum
to maximum light within the photometric uncertainties of a few \% (see
Figures~\ref{fig:phaseqgsingle} and \ref{fig:phaseqgdouble}).  This is
again consistent with a lightcurve that is produced by an elongated
shape, rather than by albedo variations.  The colors ($B-V=1.00\pm
0.04$, $V-R=0.60\pm 0.02$) show that 2001 QG$_{298}$ is red and
similar to the mean values ($B-V=0.98\pm 0.04$, $V-R=0.61\pm 0.02$, 28
objects) for KBOs as a group (Jewitt and Luu 2001).

The absolute magnitude of a Solar System object, $m_{R}(1,1,0)$, is
the hypothetical magnitude the object would have if it where at
heliocentric ($R$) and geocentric ($\Delta$) distances of 1 AU and had
a phase angle ($\alpha$) of 0 degrees.  We use the relation $m_{R}(1,1,0) = m_{R} -
5\mbox{log}(R\Delta) - \beta \alpha$ to find the absolute magnitude by
correcting for the geometrical and phase angle effects in the 2001
QG$_{298}$ observations.  Here $m_{R}$ is the apparent red magnitude
of the object and $\beta$ is the phase function.  Using the nominal
value of $\beta = 0.16$ magnitudes per degree for KBOs at low phase
angles (Sheppard \& Jewitt 2002; Sheppard \& Jewitt 2004) and data
from Table~1 we find that 2001 QG$_{298}$ has $m_{R}(1,1,0) = 6.28\pm
0.02$ at maximum light and $m_{R}(1,1,0) = 7.42\pm 0.02$ magnitudes at
minimum light.  If attributed to a rotational variation of the
cross-section, this corresponds to a ratio of maximum to minimum areas
of 2.85:1.

The effective radius of an object can be calculated using the relation
$m_{R}(1,1,0) = m_{\odot} -
2.5\mbox{log}\left[p_{R}r_{e}^{2}/2.25\times 10^{16}\right]$ where
$m_{\odot}$ is the apparent red magnitude of the sun ($-27.1$),
$p_{R}$ is the red geometric albedo and $r_{e}$ (km) is the effective
circular radius of the object.  If we assume an albedo of 0.04 (0.10)
this corresponds to effective circular radii at maximum and minimum
light of about 158 (100) km and 94 (59) km, respectively.  At the mean
absolute magnitude of 6.85 mag, the effective circular radius is 122
(77) km.

\section{Analysis}

Only three other objects in the Solar System larger than 25 km in
radius are known to have lightcurve ranges $> 1.0$ magnitude
(Table~4).  Following Jewitt and Sheppard (2002) we discuss three
possible models of rotational variation to try to compare the objects
from Table~4 with 2001 QG$_{298}$.

\subsection{Albedo Variation}
On asteroids, albedo variations contribute brightness variations that
are usually less than about $10\% - 20\%$ (Degewij, Tedesco \& Zellner
1979).  Rotationally correlated color variations may be seen if the
albedo variations are large since materials with markedly different
albedos may differ compositionally.  As seen in Table~4, Saturn's
satellite Iapetus is the only object in which variations $\ge$1
mag. are explained through albedo.  The large albedo contrast on
Iapetus is likely a special consequence of its synchronous rotation
and the anisotropic impact of material trapped in orbit about Saturn
onto its leading hemisphere (Cook \& Franklin 1970).  Iapetus shows
clear rotational color variations ($\Delta(B-V) \sim$ 0.1 mag.) that
are correlated with the rotational albedo variations (Millis 1977) and
which would be detected in 2001 QG$_{298}$ given the quality of our
data.  The special circumstance of Iapetus is without obvious analogy
in the Kuiper Belt and we do not believe that it is a good model for
the extreme lightcurve of 2001 QG$_{298}$.

Pluto shows a much smaller variation (about 0.3 magnitudes) thought to
be caused by albedo structure (Buie, Tholen \& Wasserman 1997).  Pluto
is so large that it can sustain an atmosphere which may contribute to
amplifying its lightcurve range by allowing surface frosts to condense
on brighter (cooler) spots.  Thus brighter spots grow brighter while
darker (hotter) spots grow darker through the sublimation of ices.
This positive feedback mechanism requires an atmosphere and is
unlikely to be relevant on a KBO as small as 2001 QG$_{298}$.

While we cannot absolutely exclude surface markings as the dominant
cause of 2001 QG$_{298}$'s large rotational brightness variation, we
are highly skeptical of this explanation.  We measure no color
variation with rotation, there appear to be two distinct minima and
the range is so large as to be beyond reasonable explanation from
albedo alone.

\subsection{Aspherical Shape}
Since surface markings are most likely not the cause of the lightcurve, the observed 
photometric variations are probably caused by changes in the
projected cross-section of an elongated body in rotation about its
minor axis.  The rotation period of an elongated object should be
twice the single-peaked lightcurve period because of the projection of
both long axes (2 maxima) and short axes (2 minima) during one full
rotation.  If the body is elongated, we can use the ratio of maximum
to minimum brightness to determine the projection of the body shape
into the plane of the sky.  The rotational brightness range of a
triaxial object with semiaxes $a \geq b \geq c$ in rotation about the
$c$ axis and viewed equatorially is

\begin{equation}
\Delta m=2.5\mbox{log}\left(\frac{a}{b}\right)
\label{eq:elong}
\end{equation}

where $\Delta m$ is expressed in magnitudes.  This gives a lower limit
to $a/b$ because of the effects of projection.  Using $\Delta m =
1.14$ for 2001 QG$_{298}$, we find the lower limit is $a/b=2.85$.
This corresponds to $a=267$ and $b=94$ km for the geometric albedo
0.04 case and $a=169$ and $b=59$ km for an albedo of 0.10.

It is possible that 2001 QG$_{298}$ is elongated and able to resist
gravitational compression into a spherical shape by virtue of its
intrinsic compressive strength.  However, observations of asteroids in
the main-belt suggest that only the smallest ($\sim$0.1 km sized)
asteroids are in possession of a tensile strength sufficient to resist
rotational deformation (Pravec, Harris \& Michalowski 2003).
Observations of both asteroids and planetary satellites suggest that
many objects with radii $\ge$ 50 to 75 km have shapes controlled by
self-gravity, not by material strength (Farinella 1987; Farinella \&
Zappala 1997).  The widely accepted explanation is that these bodies
are internally weak because they have been fractured by numerous past
impacts.  This explanation is also plausible in the Kuiper Belt, where
models attest to a harsh collisional environment at early times
(e.g. Davis \& Farinella 1997).  We feel that the extraordinarily
large amplitude of 2001 QG$_{298}$ is unlikely to be caused by
elongation of the object sustained by its own material strength,
although we cannot rule out this possibility.

Structurally weak bodies are susceptible to rotational
deformation. The 1000-km scale KBO (20000) Varuna (rotation period
6.3442 $\pm$ 0.0002 hr and lightcurve range 0.42 $\pm$ 0.02 mag) is
the best current example in the Kuiper Belt (Jewitt and Sheppard
2002).  In the main asteroid belt, 216 Kleopatra has a very short
period (5.385 hr) and large lightcurve range (1.18 mag., corresponding
to axis ratio $\sim$2.95:1 and dimensions $\sim$ 217 $\times$ 94 km,
Table~4).  Kleopatra has been observed to be a highly elongated body
through radar and high resolution imaging and the most likely
explanation is that 216 Kleopatra is rotationally deformed (Leone et
al. 1984; Ostro et al. 2000; Hestroffer et al. 2002; Washabaugh \&
Scheeres 2002).  Is rotational elongation a viable model for 2001
QG$_{298}$?

The critical rotation period ($T_{crit}$) at which centripetal
acceleration equals gravitational acceleration towards the center of a
rotating spherical object is

\begin{equation}
T_{crit} = \left(\frac{3\pi }{G \rho}\right)^{1/2}   \label{eq:equil}
\end{equation}

where $G$ is the gravitational constant and $\rho$ is the density of
the object.  With $\rho$ = $1000$ kg m$^{-3}$ the critical period is
about 3.3 hr.  Even at longer periods, real bodies will suffer
centripetal deformation into triaxial aspherical shapes which depend on their density, angular momentum and
material strength.  The limiting equilibrium shapes of rotating
strengthless fluid bodies have been well studied by Chandrasekhar
(1987) and a detailed discussion in the context of the KBOs can be
found in Jewitt and Sheppard (2002).  We briefly mention here that
triaxial "Jacobi" ellipsoids with large angular momenta are
rotationally elongated and generate lightcurves with substantial
ranges when viewed equatorially.

Leone et al. (1984) have analyzed rotational equilibrium
configurations of strengthless asteroids in detail (see
Figure~\ref{fig:kboperamp}). They show that the maximum photometric
range of a rotational ellipsoid is 0.9 mag: more elongated objects
are unstable to rotational fission.  The 1.14 mag photometric range of
2001 QG$_{298}$ exceeds this limit.  In addition, the 13.7744 hr
(two-peaked) rotation period is much too long to cause significant
elongation for any plausible bulk density
(Figure~\ref{fig:kboperamp}).  For these reasons we do not believe
that 2001 QG$_{298}$ is a single rotationally distorted object.

\subsection{Binary Configurations}

A third possible explanation for the extreme lightcurve of 2001
QG$_{298}$ is that this is an eclipsing binary.  A wide separation
(sum of the orbital semi-major axes much larger than the sum of the
component radii) is unlikely because such a system would generate a
distinctive ``notched'' lightcurve that is unlike the lightcurve of
2001 QG$_{298}$.  In addition, a wide separation would require
unreasonably high bulk density of the components in order to generate
the measured rotational period.  If 2001 QG$_{298}$ is a binary then
the components must be close or in contact.  We next consider the
limiting case of a contact binary.

The axis ratio of a contact binary consisting of equal spheres is $a/b
= 2$, corresponding to a lightcurve range $\Delta m$ = 0.75
magnitudes, as seen from the rotational equator.  At the average
viewing angle $\theta = 60$ degrees we would expect $\Delta m$ = 0.45
mag.  The rotational variation of 2001 QG$_{298}$ is too large to be
explained as a contact binary consisting of two equal spheres.
However, close binary components of low strength should be elongated
by mutual tidal forces, giving a larger lightcurve range than possible
in the case of equal spheres (Leone et al. 1984).  The latter authors
find that the maximum range for a tidally distorted nearly contact
binary is 1.2 magnitudes, compatible with the 1.14 mag. range of 2001
QG$_{298}$ (Figure~\ref{fig:kboperamp}). The contact binary hypothesis
is the likely explanation of 624 Hektor's lightcurve (Hartmann \&
Cruikshank 1978; Weidenschilling 1980; Leone et al. 1984) and could
also explain 216 Kleopatra's lightcurve (Leone et al. 1984; Ostro et
al. 2000; Hestroffer et al. 2002).

We suggest that the relatively long double-peaked period ($13.7744 \pm
0.0004$ hr) and large photometric range ($1.14\pm 0.04$ magnitudes) of
2001 QG$_{298}$'s lightcurve are best understood if the body is a
contact binary or nearly contact binary viewed from an approximately
equatorial perspective.  The large range suggests that the components
are of similar size and are distorted by their mutual tidal
interactions.  Using the calculations from Leone et al. (1984), who
take into account the mutual deformation of close, strengthless binary
components, we find the density of these objects must be $\sim$1000 kg
m$^{-3}$ in order to remain bound in a binary system separated by the
Roche radius (which is just over twice the component radius).  If we
assume that the albedo of both objects is 0.04, the effective radius
of each component is about 95 km as found above.  Using this
information we find from Kepler's third law that if the components are
separated, they would be about 300 km apart.  This separation as seen
on the sky ($0.01 \arcsec$) is small enough to have escaped resolution
with current technology.

Further, we point out that the maximum of the lightcurve of 2001
QG$_{298}$ is more nearly ``U'' shaped (or flattened) than is the
``V'' shaped minimum (Figure~\ref{fig:phaseqgdouble}).  This is also
true for 624 Hektor (Dunlap \& Gehrels 1969) and may be a
distinguishing, though not unique, signature of a contact or nearly
contact binary (Zappala 1984; Leone et al. 1984; Cellino et al. 1985).
In comparison, (20000) Varuna, which is probably not a contact binary
(see below and Jewitt \& Sheppard 2002), does not show significant
differences in the curvature of the lightcurve maxima and minima.

In short, while we cannot prove that 2001 QG$_{298}$ is a contact
binary, we find by elimination of other possibilities that this is the
most convincing explanation of its lightcurve.

\subsection{Fraction of Contact Binaries in the Kuiper Belt}

The distribution of measured lightcurve properties is shown in
Figure~\ref{fig:kboperamp} (adapted from Figure~4 of Leone et
al. (1984)).  There, Region A corresponds to the low rotational range
objects (of any period) in which the variability can be plausibly
associated with surface albedo markings.  Region B corresponds to the
rotationally deformed Jacobi ellipsoids while Region C marks the domain of
the close binary objects.  Plotted in
the Figure are the lightcurve periods and ranges for KBOs from the
HKBVP (Jewitt \& Sheppard 2002; Sheppard \& Jewitt 2002; Sheppard \&
Jewitt 2004).  We also show large main belt asteroids (data from
http://cfa-www.harvard.edu/iau/lists/LightcurveDat.html updated by
A. Harris and B. Warner and based on Lagerkvist, Harris \& Zappala
1989).  Once again we note that the measured KBO ranges should, in
most cases, be regarded as lower limits to the range because of the
possible effects of projection into the plane of the sky.

Of the 34 KBOs in our sample, five fall into Region C in
Figure~\ref{fig:kboperamp}.  Of these, 2001 QG$_{298}$ is by far the
best candidate for being a contact or nearly contact binary system
since it alone has a range between the $\Delta m_R \sim$ 0.9
mag. limit for a single rotational equilibrium ellipsoid and the
$\Delta m_R \sim $1.2 mag. limit for a mutually distorted close binary
(Table~5).  It is also rotating too slowly to be
substantially distorted by its own spin (Figure~\ref{fig:kboperamp}).
Both (33128) 1998 BU$_{48}$ and 2000 GN$_{171}$ are good candidates
which have large photometric ranges and relatively slow periods.  KBOs
(26308) 1998 SM$_{165}$ and (32929) 1995 QY$_{9}$ could be
rotationally deformed ellipsoids, but their relatively slow rotations
would require densities much smaller than that of water, a prospect
which we consider unlikely.

We next ask what might be the abundance of contact or close binaries
in the Kuiper Belt.  As a first estimate we assume that we have
detected one such object (2001 QG$_{298}$) in a sample of 34 KBOs
observed with adequate time resolution.  The answer depends on the
magnitude of the correction for projection effects caused by the
orientation of the rotation vector with respect to the line of sight.
This correction is intrinsically uncertain, since it depends on
unknowns such as the scattering function of the surface materials of
the KBO as well as on the detailed shape.  We adopt two crude
approximations that should give the projection correction at least to
within a factor of a few.

First, we represent the elongated shape of the KBO by a rectangular
block with dimensions a $>$ b = c.  The lightcurve range varies with
angle from the equator, $\theta$, in this approximation as

\begin{equation}
\Delta m = 2.5 log\left[\frac{1 + tan\theta}{\frac{b}{a}+ tan\theta}\right].
\end{equation}

For the limiting case of a highly distorted contact binary with
$\Delta m$ = 1.2 mag. at $\theta$ = 0$^\circ$, Eq. (3) gives $a/b$ =
3.  We next assume that the range must fall in the range 0.9 $\le
\Delta m \le$ 1.2 mag. in order for us to make an assignment of likely
binary structure (Figure~\ref{fig:kbomaghist}).  As noted above, only
2001 QG$_{298}$ satisfies this condition amongst the known objects.
We find, from Eq. (3) with $a/b$ = 3, that $\Delta m$ = 0.9 mag is
reached at $\theta$ = 10$^\circ$.  The probability that Earth would
lie within 10$^{\circ}$ of the equator of a set of randomly oriented
KBOs is $P(\theta \le 10)$ = 0.17.  Therefore, the detection of 1 KBO
with 0.9 $\le \Delta m \le$ 1.2 mag implies that the fractional
abundance of similarly elongated objects is $f \sim 1/(34P) \sim$
17\%.

As a separate check on this estimate, we next represent the object as
an ellipsoid, again with axes a $>$ b = c.  The photometric range when
viewed at an angle $\theta$ from the rotational equator is given by

\begin{equation}
\Delta m = 2.5 log(a/b) - 1.25 log \left[ \left[\left(\frac{a}{b}\right)^2 -1\right]sin^2\theta + 1\right]
\end{equation}

Substituting $a/b$ = 3, the range predicted by Eq. (4) falls to 0.9
mag at $\theta \sim$ 17$^\circ$.  Given a random distribution of the
spin vectors, the probability that Earth would lie within 17$^{\circ}$
of the equator is $P(\theta \le 17)$ = 0.29.  Therefore, the detection
of 1 KBO with a range between 0.9 and 1.2 mag in a sample of 34
objects implies, in this approximation, a fractional abundance of
similarly elongated objects near $f \sim 1/(34P) \sim$ 10\%.

Given the crudity of the model, the agreement between projection
factors from Eqs. (3) and (4) is encouraging.  Together, the data and
the projection factors suggest that in our sample of 34 KBOs, perhaps
3 to 6 objects are as elongated as 2001 QG$_{298}$ but only 2001
QG$_{298}$ is viewed from a sufficiently equatorial perspective that
the lightcurve is distinct.  This is consistent with Figure 5, which
shows that $5$ of $34$ KBOs ($15\%$) from the HKBVP occupy Region C of
the period-range diagram.  Our estimate is very crude and is also a
lower limit to the true binary fraction because close binaries with
components of unequal size will not satisfy the 0.9 $\le \Delta m \le$
1.2 mag. criterion for detection.  The key point is that the data are
consistent with a substantial close binary fraction in the Kuiper Belt .

Figure~\ref{fig:kboperamp} also shows that there are no large main-belt
asteroids (radii $\ge 100$ km) in Region C, which is where similar
sized component contact binaries are expected to be.  To date, no
examples of large binary main-belt asteroids with similar sized
components have been found, even though the main belt has been
extensively searched for binarity (see Margot 2002 and references
therein).  The main-belt asteroids may have had a collisional history
significantly different from that of the KBOs.

The contact binary interpretation of the 2001 QG$_{298}$ lightcurve is
clearly non-unique.  Indeed, firm proof of the existence of contact
binaries will be as difficult to establish in the Kuiper Belt as it
has been in closer, brighter populations of small bodies.
Nevertheless, the data are compatible with a high abundance of such
objects.  It is interesting to speculate about how such objects could
form in abundance.  One model of the formation and long term evolution
of wide binaries predicts that such objects could be driven together
by dynamical friction or three-body interactions (Goldreich et
al. 2002).  Objects like 2001 QG$_{298}$ would be naturally produced
by such a mechanism.

\section{Summary}

Kuiper Belt Object 2001 QG$_{298}$ has the most extreme lightcurve of
any of the 34 objects so far observed in the Hawaii Kuiper Belt
Variability Project.

1.  The double-peaked lightcurve period is $13.7744 \pm 0.0004$ hr and
    peak-to-peak range is $1.14 \pm 0.04$ mag.  Only two other minor
    planets with radii $\ge$ 25 km (624 Hektor and 216 Kleopatra) and
    one planetary satellite (Iapetus) are known to show rotational
    photometric variation greater than 1 mag.

2.  The absolute red magnitude is $m_R$(1,1,0) = 6.28 at maximum light
    and 7.42 mag. at minimum light.  With an assumed geometric albedo
    of 0.04 (0.10) we derive effective circular radii at maximum and
    minimum light of 158 (100) and 94 (59) km, respectively.

3.  No variation in the BVR colors between maximum and minimum light was
    detected to within photometric uncertainties of a few percent.
    
4. The large photometric range, differences in the lightcurve minima, and
   long period of 2001 QG$_{298}$ are consistent with and strongly
   suggest that this object is a contact or nearly contact binary,
   viewed equatorially.

5.  If 2001 QG$_{298}$ is a contact binary with similarly sized
    components, then we conclude that such objects constitute at least
    10\% to 20\% of the Kuiper Belt population at large sizes.

\section*{Acknowledgments}

We thank John Tonry and Andrew Pickles for help with the OPTIC camera
and the remote observing system on the University of Hawaii 2.2 meter
telescope.  We also thank Henry Hsieh for observational assistance and
Jane Luu for comments on the manuscript.  This work was supported by a
grant to D.J. from the NASA Origins Program.

\newpage

\begin{figure}
\caption{The phase dispersion minimization (PDM) plot for 2001
QG$_{298}$.  A smaller theta corresponds to a better fit.  Best fits
from this plot are the 6.8872 hour single-peaked fit and the 13.7744
hour double-peaked fit.  Both are flanked by alias periods.}
\label{fig:pdmqg} 
\end{figure}

\begin{figure}
\caption{Phased data from all the observations in 2002 and 2003 of
2001 QG$_{298}$.  The period has been phased to 6.8872 hr which is the
best fit single-peaked period.  Filled colored symbols are data taken
in the B-band (blue), V-band (green) and R-band (red) at the Keck I
telescope on UT August 30.  All other symbols are R-band data from the
various nights of observations at UH 2.2 m telescope.  The $B$ and $V$
points have been shifted according to their color differences from the
R-band ($V-R = 0.60$ and $B-V = 1.00$).  No color variation is seen
between maximum and minimum light.  The uncertainty on each
photometric observation is $\pm 0.03$ mag.}
\label{fig:phaseqgsingle} 
\end{figure}

\begin{figure}
\caption{Phased data from all the observations in 2002 and 2003 of
2001 QG$_{298}$.  The period has been phased to 13.7744 hr which is
the best fit double-peaked period.  Filled colored symbols are data
taken in the B-band (blue), V-band (green) and R-band (red) at the
Keck I telescope on UT August 30.  All other symbols are R-band data
from the various nights of observations at UH 2.2 m telescope.  The
$B$ and $V$ points have been shifted according to their color
differences from the R-band ($V-R = 0.60$ and $B-V = 1.00$).  There
appears to be two distinct minima.  The minima appear to be more
``notched'' compared to the flatter maxima.  No color variation is
seen between maximum and minimum light.  The uncertainty for each
photometric observation is $\pm 0.03$ mag.}
\label{fig:phaseqgdouble} 
\end{figure}

\begin{figure}
\caption{A closer view of the phase dispersion minimization (PDM) plot
for 2001 QG$_{298}$ around the double-peaked period at 13.7744 hr.
The best fit is flanked by aliases from separation of the 3 data sets
obtained for this object.  Only the center PDM peak fits the data once
it is phased together.}
\label{fig:pdmqgsmall} 
\end{figure}

\begin{figure}
\caption{ This Figure is a modification of Figure 4 from Leone et
al. (1984).  We here show the rotation periods and photometric ranges
of known KBO lightcurves and the larger asteroids.  The Regions are
defined as A) The range of the lightcurve could be equally caused by
albedo, elongation or binarity B) The lightcurve range is most likely
caused by rotational elongation C) The lightcurve range is most likely
caused by binarity of the object. Stars denote KBOs, Circles denote
main-belt asteroids (radii $\ge 100$ km) and Squares denote the Trojan
624 Hektor and the main-belt asteroid 216 Kleopatra. Objects just to
the left of Region B would have densities significantly less than 1000
kg m$^{-3}$ in order to be elongated from rotational angular momentum.
Binary objects are not expected to have photometric ranges above 1.2
magnitudes.  The 23 KBOs which have photometric ranges below our
photometric uncertainties ($\sim 0.1$ mag) in our Hawaii survey have
not been plotted since their periods are unknown.  These objects would
all fall into Region A.  The asteroids have been plotted at their
expected mean projected viewing angle of 60 degrees in order to more
directly compare to the KBOs of unknown projection angle.}
\label{fig:kboperamp} 
\end{figure}

\begin{figure}
\caption{A histogram of known KBOs photometric ranges.  There is a
break in known photometric ranges starting around 0.25 magnitudes.
The Regions are defined as 1) The lightcurve range could be dominated
by albedo, elongation or binarity 2) The lightcurve is likely
dominated from rotational elongation or binarity 3) The lightcurve is
likely caused by binarity.  Data is from our Hawaii Kuiper Belt object
variability project (Sheppard \& Jewitt 2002 and Sheppard and Jewitt
2004).}
\label{fig:kbomaghist} 
\end{figure}

\end{document}